


\hsize=6.0truein
\vsize=8.5truein
\voffset=0.25truein
\hoffset=0.1875truein
\tolerance=1000
\hyphenpenalty=500
\def\monthintext{\ifcase\month\or January\or February\or
   March\or April\or May\or June\or July\or August\or
   September\or October\or November\or December\fi}


\font\tenrm=cmr10 scaled \magstep1   \font\tenbf=cmbx10 scaled \magstep1
\font\sevenrm=cmr7 scaled \magstep1  
\font\fiverm=cmr5 scaled \magstep1   

\font\teni=cmmi10 scaled \magstep1   \font\tensy=cmsy10 scaled \magstep1
\font\seveni=cmmi7 scaled \magstep1  \font\sevensy=cmsy7 scaled \magstep1
\font\fivei=cmmi5 scaled \magstep1   \font\fivesy=cmsy5 scaled \magstep1

\font\tentt=cmtt10 scaled \magstep1
\font\tenit=cmti10 scaled \magstep1
\font\tensl=cmsl10 scaled \magstep1

\def\twelvepoint{\def\rm{\fam0\tenrm}
   \textfont0=\tenrm \scriptfont0=\sevenrm \scriptscriptfont0=\fiverm
   \textfont1=\teni  \scriptfont1=\seveni  \scriptscriptfont1=\fivei
   \textfont2=\tensy \scriptfont2=\sevensy \scriptscriptfont2=\fivesy
   \textfont\itfam=\tenit \def\it{\fam\itfam\tenit}
   \textfont\ttfam=\tentt \def\tt{\fam\ttfam\tentt}
   \textfont\bffam=\tenbf \def\bf{\fam\bffam\tenbf}
   \textfont\slfam=\tensl \def\sl{\fam\slfam\tensl} \rm
   \hfuzz=1pt\vfuzz=1pt
   \setbox\strutbox=\hbox{\vrule height 10.2pt depth 4.2pt width 0pt}
   \parindent=24pt\parskip=1.2pt plus 1.2pt
   \topskip=12pt\maxdepth=4.8pt\jot=3.6pt
   \normalbaselineskip=14.4pt\normallineskip=1.2pt
   \normallineskiplimit=0pt\normalbaselines
   \abovedisplayskip=13pt plus 3.6pt minus 5.8pt
   \belowdisplayskip=13pt plus 3.6pt minus 5.8pt
   \abovedisplayshortskip=-1.4pt plus 3.6pt
   \belowdisplayshortskip=13pt plus 3.6pt minus 3.6pt
   \topskip=12pt \splittopskip=12pt
   \scriptspace=0.6pt\nulldelimiterspace=1.44pt\delimitershortfall=6pt
   \thinmuskip=3.6mu\medmuskip=3.6mu plus 1.2mu minus 1.2mu
   \thickmuskip=4mu plus 2mu minus 1mu
   \smallskipamount=3.6pt plus 1.2pt minus 1.2pt
   \medskipamount=7.2pt plus 2.4pt minus 2.4pt
   \bigskipamount=14.4pt plus 4.8pt minus 4.8pt}

\twelvepoint



\font\titlerm=cmr10 scaled \magstep3
\font\titlerms=cmr10 scaled \magstep1 
\font\titlei=cmmi10 scaled \magstep3  
\font\titleis=cmmi10 scaled \magstep1 
\font\titlesy=cmsy10 scaled \magstep3   
\font\titlesys=cmsy10 scaled \magstep1  
\font\titleit=cmti10 scaled \magstep3   
\skewchar\titlei='177 \skewchar\titleis='177 
\skewchar\titlesy='60 \skewchar\titlesys='60 

\def\titlefont{\def\rm{\fam0\titlerm}
   \textfont0=\titlerm \scriptfont0=\titlerms 
   \textfont1=\titlei  \scriptfont1=\titleis  
   \textfont2=\titlesy \scriptfont2=\titlesys 
   \textfont\itfam=\titleit \def\it{\fam\itfam\titleit} \rm}


\def\preprint#1{\baselineskip=19pt plus 0.2pt minus 0.2pt \pageno=0
   \begingroup
   \nopagenumbers\parindent=0pt\baselineskip=14.4pt\rightline{#1}}
\def\title#1{
   \vskip 0.9in plus 0.45in
   \centerline{\titlefont #1}}
\def\secondtitle#1{}
\def\author#1#2#3{\vskip 0.9in plus 0.45in
   \centerline{{\bf #1}\myfoot{#2}{#3}}\vskip 0.12in plus 0.02in}
\def\secondauthor#1#2#3{}
\def\addressline#1{\centerline{#1}}
\def\abstract{\vskip 0.7in plus 0.35in
     \centerline{\bf Abstract}
     \smallskip}
\def\finishtitlepage#1{\vskip 0.8in plus 0.4in
   \leftline{#1}\supereject\endgroup}

\def\date#1{\finishtitlepage{#1}}

\def\nolabels{\def\eqnlabel##1{}\def\eqlabel##1{}\def\figlabel##1{}%
     \def\reflabel##1{}}
\def\writelabels{\def\eqnlabel##1{%
     {\escapechar=` \hfill\rlap{\hskip.11in\string##1}}}%
     \def\eqlabel##1{{\escapechar=` \rlap{\hskip.11in\string##1}}}%
     \def\figlabel##1{\noexpand\llap{\string\string\string##1\hskip.66in}}%
     \def\reflabel##1{\noexpand\llap{\string\string\string##1\hskip.37in}}}
\nolabels


\global\newcount\secno \global\secno=0
\global\newcount\meqno \global\meqno=1

\def\newsec#1{\global\advance\secno by1
   \xdef\secsym{\the\secno.}
   \global\meqno=1\bigbreak\medskip
   \noindent{\bf\the\secno. #1}\par\nobreak\smallskip\nobreak\noindent}
\xdef\secsym{}

\def\appendix#1#2{\global\meqno=1\xdef\secsym{\hbox{#1.}}\bigbreak\medskip
\noindent{\bf Appendix #1. #2}\par\nobreak\smallskip\nobreak\noindent}


\def\eqnn#1{\xdef #1{(\secsym\the\meqno)}%
     \global\advance\meqno by1\eqnlabel#1}
\def\eqna#1{\xdef #1##1{\hbox{$(\secsym\the\meqno##1)$}}%
     \global\advance\meqno by1\eqnlabel{#1$\{\}$}}
\def\eqn#1#2{\xdef #1{(\secsym\the\meqno)}\global\advance\meqno by1%
     $$#2\eqno#1\eqlabel#1$$}


\def\myfoot#1#2{{\baselineskip=14.4pt plus 0.3pt\footnote{#1}{#2}}}
\global\newcount\ftno \global\ftno=1
\def\foot#1{{\baselineskip=14.4pt plus 0.3pt\footnote{$^{\the\ftno}$}{#1}}%
     \global\advance\ftno by1}


\global\newcount\refno \global\refno=1
\newwrite\rfile

\def\ref{[\the\refno]\nref}
\def\nref#1{\xdef#1{[\the\refno]}\ifnum\refno=1\immediate
 \openout\rfile=refs.tmp\fi\global\advance\refno by1\chardef\wfile=\rfile
 \immediate\write\rfile{\noexpand\item{#1\ }\reflabel{#1}\pctsign}\findarg}
\def\findarg#1#{\begingroup\obeylines\newlinechar=`\^^M\passarg}
{\obeylines\gdef\passarg#1{\writeline\relax #1^^M\hbox{}^^M}%
\gdef\writeline#1^^M{\expandafter\toks0\expandafter{\striprelax #1}%
\edef\next{\the\toks0}\ifx\next\null\let\next=\endgroup\else\ifx\next\empty%
\else\immediate\write\wfile{\the\toks0}\fi\let\next=\writeline\fi\next\relax}}
     {\catcode`\%=12\xdef\pctsign{
\def\striprelax#1{}

\def\semi{;\hfil\break}
\def\addref#1{\immediate\write\rfile{\noexpand\item{}#1}} 

\def\listrefs{\vfill\eject\immediate\closeout\rfile
   \centerline{{\bf References}}\bigskip{\frenchspacing%
   \catcode`\@=11\escapechar=` %
   \input refs.tmp\vfill\eject}\nonfrenchspacing}

\def\startrefs#1{\immediate\openout\rfile=refs.tmp\refno=#1}


\global\newcount\figno \global\figno=1
\newwrite\ffile
\def\fig{\the\figno\nfig}
\def\nfig#1{\xdef#1{\the\figno}\ifnum\figno=1\immediate
     \openout\ffile=figs.tmp\fi\global\advance\figno by1\chardef\wfile=\ffile
     \immediate\write\ffile{\medskip\noexpand\item{Fig.\ #1:\ }%
     \figlabel{#1}\pctsign}\findarg}

\def\listfigs{\vfill\eject\immediate\closeout\ffile{\parindent48pt
     \baselineskip16.8pt\centerline{{\bf Figure Captions}}\medskip
     \escapechar=` \input figs.tmp\vfill\eject}}


\def\letter{\raggedright\parindent=0pt}
\def\endmode{}
\def\longindent{\parindent=3.25truein\obeylines\parskip=0pt}
\def\letterhead{\null\vfil\begingroup
   \parindent=3.25truein\obeylines
   \def\endmode{\medskip\endgroup}}

\def\sendingaddress{\endmode\begingroup
   \parindent=0pt\obeylines\def\endmode{\medskip\endgroup}}

\def\salutation{\endmode\begingroup
   \parindent=0pt\obeylines\def\endmode{\medskip\endgroup}}

\def\body{\endmode\begingroup\parskip=\smallskipamount
   \def\endmode{\medskip\endgroup}}

\def\closing{\endmode\begingroup\longindent
   \def\endmode{\endgroup}}

\def\signed{\endmode\begingroup\longindent\vskip0.8truein
   \def\endmode{\endgroup}}

\def\endofletter{\endmode \ifnum\pageno=1 \nopagenumbers\fi
     \vfil\vfil\eject\end}


\def\noblackbox{\overfullrule=0pt}
\def\inv{^{\raise.18ex\hbox{${\scriptscriptstyle -}$}\kern-.06em 1}}
\def\dup{^{\vphantom{1}}}
\def\Dsl{\,\raise.18ex\hbox{/}\mkern-16.2mu D} 
\def\dsl{\raise.18ex\hbox{/}\kern-.68em\partial}
\def\slash#1{\raise.18ex\hbox{/}\kern-.68em #1}
\def\lspace{}
\def\lbspace{}
\def\boxeqn#1{\vcenter{\vbox{\hrule\hbox{\vrule\kern3.6pt\vbox{\kern3.6pt
     \hbox{${\displaystyle #1}$}\kern3.6pt}\kern3.6pt\vrule}\hrule}}}
\def\mbox#1#2{\vcenter{\hrule \hbox{\vrule height#2.4in
     \kern#1.2in \vrule} \hrule}}  
\def\bar{\overline}
\def\e#1{{\rm e}^{\textstyle#1}}
\def\del{\partial}
\def\curly#1{{\hbox{{$\cal #1$}}}}
\def\curlyD{\hbox{{$\cal D$}}}
\def\curlyL{\hbox{{$\cal L$}}}
\def\vev#1{\langle #1 \rangle}
\def\psibar{\overline\psi}
\def\lform{\hbox{$\sqcup$}\llap{\hbox{$\sqcap$}}}
\def\darr#1{\raise1.8ex\hbox{$\leftrightarrow$}\mkern-19.8mu #1}
\def\half{{\textstyle{1\over2}}} 
\def\roughly#1{\ \lower1.5ex\hbox{$\sim$}\mkern-22.8mu #1\,}
\def\MSbar{$\bar{{\rm MS}}$}
\hyphenation{di-men-sion di-men-sion-al di-men-sion-al-ly}


\def\third{{\textstyle{1\over 3}}}
\def\qutr{{\textstyle{1\over 4}}}
\def\hf{\half}
\def\nonp{non-perturbative}
\def\hmm{hermitian matrix model}
\def\integ#1#2#3{\int_{#1}^{#2}\!\!\! d#3\ }
\def\h{$\cal H$}
\def\O{{\cal O}}
\def\R{{\cal R}}
\def\rline{{\rm I}\!{\rm R}} 
\def\pq{$[P,Q]=1$ }
\def\pqq{$[{\tilde P},Q]=Q$ }
\def\nuke{Nucl.Phys.}
\def\pl{Phys.Lett.}
\def\re{\rho_R}
\def\im{\rho_I}
\def\s{$\sigma$}
\def\ob{$\O_B$}
\def\trho{\upsilon}
\def\NP#1{Nucl. Phys.\ {\bf #1}\ }
\def\PL#1{Phys. Lett.\ {\bf #1}\ }
\def\CMP#1{Comm. Math. Phys.\ {\bf #1}\ }
\def\LNC#1{Lett. Nuovo Cimento\ {\bf #1}\ }
\def\NC#1{Nuovo Cimento\ {\bf #1}\ }
\def\CQG#1{Class. Quantum Grav.\ {\bf #1}\ }
\def\PR#1{Phys. Rev\ {\bf #1}\ }
\def\PREP#1{Phys. Rep\ {\bf #1}\ }
\def\PRL#1{Phys. Rev. Lett\ {\bf #1}\ }
\def\JMP#1{J. Math. Phys\ {\bf #1}\ }
\def\SAM#1{Stud. Appl. Math\ {\bf #1}\ }
\def\MPL#1{Mod. Phys. Lett\ {\bf #1}\ }
\def\PA#1{Physica\ {\bf #1}\ }
\def\PP#1{Princeton preprint\ { #1}\ }
\def\RP#1{Rutgers preprint\ { #1}\ }
\def\sec#1{
\bigskip
\noindent{\bf #1}
\bigskip}

\preprint{SHEP 91/92--21}
\title{\vbox{
\centerline{The Boundary Cosmological Constant}
\vskip2pt\centerline{in}
\vskip2pt\centerline{Stable 2D Quantum Gravity}}}
\author{Clifford V. Johnson, Tim R. Morris and Peter L. White}{}{}
\addressline{\it Physics Department}
\addressline{\it University of Southampton}
\addressline{\it Southampton}
\addressline{\it SO9 5NH, U.K.}
\vskip -1.5cm

\abstract
We study further the r\^ole of the boundary operator $\O_B$ for macroscopic
loop length in the stable definition  of 2D quantum gravity provided by the
\pqq formulation. The KdV flows are supplemented by an additional flow with
respect  to the  boundary cosmological  constant \s.  We numerically  study
these  flows for  the $m=1$,  $2$ and  $3$ models,  solving for  the string
susceptibility  in the  presence of  $\O_B$ for  arbitrary coupling \s. The
spectrum of the Hamiltonian of the loop quantum mechanics is continuous and
bounded from below by \s. For large positive \s, the theory is dominated by
the  `universal'   $m=0$  topological  phase  present   only  in  the  \pqq
formulation.   For  large   negative  \s,   the  non--perturbative  physics
approaches that  of the \pq  definition, although there  is no path  to the
unstable solutions of the \pq $m$-even models.

\date{May 1992.}

In   certain    toy   models   of   closed    string   theory,   the   \pqq
formulation\ref\npqg{S.Dalley,  C.V.Johnson  and  T.R.Morris,  \nuke\  {\bf
B368}  (1992) 655.}  supplies the  most general  string equation compatible
with  the  KdV   flows:  \eqn\smileyi{\rho\R^2-{1\over  2}\R\R^{''}+{1\over
4}(\R^{'})^2=0.}   Here  $\R\equiv\sum_{k=0}^\infty(k+\half)t_k\R_k[\rho]$,
where  the  $\R_k$  are  the  Gel'fand--Dikii  differential  polynomials in
$\rho$. A prime denotes $\nu\partial/\partial  z$ where $\nu$ is the string
coupling. The  quantity $z=-t_0$ and the  normalisation chosen is $\R_0=2$,
which  fixes  the  other  $\R_k$   by  virtue  of  the  recursion  relation
$\R^{'}_{k+1}=\qutr\R^{'''}_k-\half\rho^{'}\R_k-\rho\R^{'}_k$    and    the
requirement that they vanish when $\rho=0$.

Setting $t_k\sim\delta_{km}$ selects the  string equation of the $(2m-1,2)$
conformal minimal model coupled to  2D quantum gravity. In particular $m=2$
is pure  2D quantum gravity.  In this  case  the parameter $z$  is the bulk
`cosmological constant', the chemical potential  for the area of the random
surfaces.  In   operator  language,  $z=-t_0$  couples   to  the  `puncture
operator', ${\cal O}_0$ in the  action. ${\cal O}_0$ is the gravitationally
dressed identity  operator of the  $(3,2)$ model and  may be thought  of as
marking  out  a  point  on  the  surface.  The  other $t_k$ couple to other
operators ${\cal O}_k$ in the theory. The function $\rho(z)$ is the `string
susceptibility',      the      two--point      function      of     $\O_0$:
$\rho=<\O_0\O_0>=-\Gamma^{''}$. $\Gamma$ is the  free energy of the theory.
The presence  of the KdV flows  in the $(2m-1,2)$ models  is most naturally
phrased             in             terms             of             $\rho$:
$\partial_{t_k}\rho\equiv<\O_k\O_0\O_0>=\R^{'}_{k+1}$. This  is a statement
about the  organisation of the  operator structure of  the theory and  flow
between all  the minimal models.  Using this relation,  the string equation
and the recursion  relation, all correlators of the  local operators $\O_k$
may be determined.

The \pq definition  of these models selects the  solutions of \smileyi\ for
the    string   susceptibility    which   satisfy    $\R=0$   everywhere.
Asymptotically,  for  the  $m$th  critical point,  we  have
 $\rho\sim  z^{1/m}$ for
$z\to\pm\infty$. This leads to the  familiar cases where for $m$-even, real
solutions  are  problematic\ref\david{F.David,  Mod.  Phys.  Lett. {\bf A5}
(1990) 1019\semi  F.David, Nucl.Phys. {\bf  B348} (1991) 507.}  and for $m$
odd,  we  have  the  BMP--type  solution\ref\bmp{E.Br\'{e}zin, E.Marinari and
G.Parisi,  \PL{B242}  (1990)  35.}.  The  \pqq  definition  of these models
selects    the   {\sl    unique}\ref\Global{C.V.Johnson,   T.R.Morris   and
A.W\"atterstam,  Southampton  preprint  SHEP  91/92--25  and G\"oteborg ITP
92--21.}\ solution of \smileyi\ with the asymptotics $\rho\sim z^{1/m}$ for
$z\to+\infty$ and $\rho\sim 0$ for $z\to-\infty$. These solutions provide a
definition for  $\rho$ which is free  from non--perturbative instabilities.
Note that  for $m$ odd that  the \pqq solution for  $\rho$ differs from the
corresponding   \pq   solution,   although   they   are   both  unique  and
well--defined.

Another distinguishing feature  of the \pqq definition is  the existence of
an  $m=0$  critical  point.  The  \pq  definition  does  not possess such a
point\ref\unitary{S.Dalley, C.V.Johnson,  T.R.Morris and  A.W\"atterstam,
Princeton preprint PUPT-1325, Southampton
preprint  SHEP  91/92--19  and  G\"oteborg  ITP  92--20.}. Setting $t_k=0$,
($k\neq 0$)  in \smileyi\ yields the  simple solution $\rho=-\nu^2/(4z^2)$.
This solution for $\rho$ does not support finite area surfaces and leads to
recurrence relations which determine all  operator products. Thus this $m=0$
critical point is a
purely  topological theory,  in addition  to the  usual $m=1$  topological
point.   This  distinguishing   feature  feeds   into  all   of  the  other
$m$--critical  models in  the \pqq   definition by  supplying them  with an
interpretation  in  the  `weak  coupling\unitary'  phase, the $z\to-\infty$
limit. In this limit, the leading  correction to the $\rho=0$ behaviour for
any  $m$ is  $\rho=-\nu^2/(4z^2)$, and  so the  universal $m=0$ topological
phase dominates the physics in this regime.

The models may be studied further  by the inclusion of macroscopic loops in
the  theory\ref\macroloops{T.Banks, M.R.Douglas,  N.Seiberg and  S.Shenker,
\PL{B238}  (1990) 279.}.  The features  of the  solution for $\rho$ receive
further attention here, for the  discussion of macroscopic loop dynamics in
these  models  reduces  to  1D  quantum  mechanics  with Hamiltonian ${\cal
H}\equiv   Q=\nu^2\partial_z^2-\rho$.  Here,   the  usual   \hmm\  inspired
prescription   for  the   expectation   value   of  the   macroscopic  loop
wavefunction,                                           \eqn\loop{<w(\ell)>
=\int_\mu^\infty<z|\e{\ell(\nu^2\partial^2_z-\rho)}|z>dz} is  adopted. For
the \pq definition, \h\ has a discrete spectrum for the $m$--even models. A
simple  exponential  behaviour  for  $<w(\ell)>$, $\e{-\epsilon_0\ell}$, is
obtained only in  the large $\ell$ limit, where  $\epsilon_0$ is the lowest
eigenvalue of \h. Meanwhile for the $m$--odd solutions the loop expectation
diverges as $\ell\to\infty$ due to the  presence of the exponential tail of
the  scaled  charge  density  in  the  associated  Dyson--gas--on--$\rline$
problem\bmp.  Here again,  the  \pqq  definition distinguishes  itself. The
$m$--critical models  all have the  same generic behaviour,  in contrast to
the \pq definition.  In considering the loop structure  of the \pqq theory,
very  simple  arguments\ref\npqga{S.Dalley,   C.V.Johnson  and  T.R.Morris,
\nuke\  {\bf B}  (Proc. Suppl.)  {\bf 25A}  (1992) 87,  Proceedings of  the
workshop on {\it Random Surfaces  and 2D Quantum Gravity,} Barcelona 10--14
June 1991.}\  yield the exponentially decreasing  behaviour for loops. This
arises naturally from  the fact that the \pqq  definition has a realisation
in terms of a Dyson gas on $\rline_+$. This will be reviewed below.

The length of loops is discussed  most naturally by the introduction of the
boundary length operator $\O_B$ to  measure it. The associated coupling will
be denoted  by $\sigma$, the  boundary cosmological constant.  The operator
$\O_B$   in  the   \pq  definition    of  these   models  was   studied  in
ref.\ref\bound{E.Martinec, G.Moore  and N.Seiberg, \PL{B263}  (1991) 190.}.
Its signature is identified by noting that $\sigma$ may be set to zero by a
redefinition of the couplings $t_k$ of the bulk operators $\O_k$ ($k<m$) in
the $m$th model\bound\ and a shift  in $\rho$. Perturbatively in the $t_k$,
$\sigma$ is  identified with the combination  $t_{m-1}/t_{m}$ for the $m$th
model.  Meanwhile $\O_B$  is a  mixture of  $\O_{m-1}$ with  all the  lower
operators   \eqn\mix{\O_B=\sum_{k=1}^{m}(k+{1\over   2})t_m\O_{m-1}+\cdots}
where  the dots  represent higher  order terms  in $t_{m-1}$.  In the  \pqq
definition  the boundary  operator and  its coupling  \s\ arise  in a  more
natural fashion than in the previous study. In ref.\npqga, the most general
string  equation in  the presence  of the  coupling $\sigma$ and compatible
with  the  KdV  flows  was derived: \eqn\smileyii{(\rho-\sigma)\R^2-{1\over
2}\R\R^{''}+{1\over 4}(\R^{'})^2=0}  which, by virtue of  the fact that the
first   derivative   of   \smileyii\   is   a   scaling   equation,  yields
\eqn\sigflow{{\partial\rho\over\partial\sigma}=-\R^{'}.}    The   equations
\smileyii\ and  \sigflow\ arise naturally in  the double scaled limit  of a
Dyson   gas   on   $\rline_+$\npqga\ref\simon{S.Dalley,   Mod. Phys. Lett.
{\bf A7} (1992) 1263.}.  The
parameter  \s\ represents  the scaled  position of  the infinite  potential
`wall' defining the $\rline_+$ topology.

The identification  of \ob\ is  most easily made\npqga\simon\  by observing
that   equations  \smileyii\   and  \sigflow\   are  invariant   under  the
infinitesimal             transformation:            \eqn\galileo{\eqalign{
&\rho\to{\tilde\rho}=\rho-\epsilon\cr
&\sigma\to{\tilde\sigma}=\sigma-\epsilon\cr                    &z\to{\tilde
z}=z-\epsilon{3\over  2}t_1\cr  &t_k\to{\tilde  t_k}=t_k+\epsilon(k+{3\over
2})t_{k+1}\cr  }} which  is a  Galilean transformation,  supplemented by  a
transformation for \s. The redundancy of \s\ is manifest here, as it may be
set  to zero  by performing  a {\sl  finite} Galilean  transformation, which
redefines all  the couplings $t_k$ and  shifts $\rho$ by $-\sigma$.  At the
same  time,  using  equation  \loop\  for  the  expectation  value  of  the
macroscopic loop  wavefunction, we see  that the resulting  shift in $\rho$
displays   the  $\e{-\sigma   l}$  loop   behaviour.  This   completes  the
identification of \s\ as the boundary length operator's coupling. Meanwhile
equation \sigflow\ yields the precise  combination of the $\O_k$ which make
up        the        boundary        operator:        \eqn\boundi{\eqalign{
{\partial\over\partial\sigma}<w(\ell)>=                                   &
\int_\mu^\infty<z|{\partial\over\partial\sigma}
\e{\ell(\nu^2\partial_z^2-\rho)}|z>dz           \cr           =           &
\ell<w(\ell)>+\sum_{k=1}^\infty(k+{1\over   2})t_k   {\partial\over\partial
t_{k-1}}<w(\ell)>    }}    from     which    we    deduce:    \eqn\boundii{
{\partial\over\partial{\tilde\sigma}}<w(\ell)>
\equiv<\O_Bw(\ell)>=\ell<w(\ell)>}                                    where
\eqn\boundiii{{\partial\over\partial\tilde\sigma}\equiv{\partial\over
\partial\sigma} -\sum_{k=1}^\infty(k+{1\over  2})t_k {\partial\over\partial
t_{k-1}}} is the insertion of the  boundary operator. We thus have the Ward
identity   \eqn\ward{<\O_B\prod_iw(\ell_i)>=\sum_j\ell_j<\prod_iw(\ell_i)>}
which is usually  referred to as the $L_{-1}$ Ward  identity. Indeed in the
language of Virasoro constraints, equations \sigflow\ and \smileyii\ may be
written  as the  familiar $L_{-1}$  and $L_0$  constraints, with a boundary
term   due  to   the  presence   of  \s:  $L_{-1}\tau=\partial_\sigma\tau$;
$L_0\tau=\sigma\partial_\sigma\tau$.   Similarly,  the   rest  of  Virasoro
constraints  appear  as  the  familiar  ones  modified  by  boundary terms:
$L_n\tau=\sigma^{n+1}\partial_\sigma\tau$. So  in the \pqq  definition, the
redundancy  of the  coupling \s\  turns into  a redundancy  of the $L_{-1}$
constraint,  as the  finite Galilean  transformation above  amounts to  the
following          automorphism          of          the         constraint
algebra\npqga\ref\pqmodels{C.V.Johnson,     T.R.Morris     and    B.Spence,
Southampton    preprint   SHEP    90/91-30.}:   ${\tilde    L}_n=\e{-\sigma
L_{-1}}L_n\e{\sigma   L_{-1}}$,   under   which   the  $L_{-1}$  constraint
disappears, and we are left with ${\tilde L}_n\tau=0$ ($n\ge 0$).

In ref.\simon, the loop equations for the \pqq definition of the $(2m-1,2)$
models were derived  and studied. Here, we turn our  attention to the study
of equations \smileyii\ and \sigflow. For $\sigma=0$ we have the previously
studied  cases: there  is a  unique pole--free  solution\foot{This has been
demonstrated for $m$=1,2 and 3, and conjectured to be true for $m>3$.}\ for
the $m$th  model with the asymptotics  $\rho\sim z^{1/m}$ for $z\to+\infty$
and  $\rho=0$   for  $z\to-\infty$.  These  solutions   have  been  studied
analytically  and numerically  elsewhere\npqg\Global\unitary. For non--zero
\s, an  asymptotic analysis of the  string equation is very  similar to the
case for $\sigma=0$. It will not  be repeated here. The leading exponential
corrections  to  the  asymptotics   $\rho=z^{1/m}$  for  $z\to+\infty$  and
$\rho=\sigma$ for $z\to-\infty$  can be calculated using a  WKB ansatz. The
coefficients  of  these  exponentials  are  all  determined by the boundary
conditions. Therefore all of the integration constants have been determined
locally, and there is at most a discrete number of solutions with the above
asymptotics. A consequence  of this is that the  only infinitesimal change
that  we can  make to  the solution  is via  KdV flows  in the $t_k$ or the
\s--flow  with equation  \sigflow. So  for any  $m$--critical model,  local
solution space is spanned by one--parameter  KdV flows to other models, and
the flow in the single parameter \s\ given by equation \sigflow. Therefore,
beginning  at any  of these  unique solutions,  any solution obtained using
these flows is also unique.

The KdV flows  between models were studied in  ref.\Global, where numerical
methods  were used  to display   intermediate flows,  the solutions  to the
interpolating string equation \smileyi. Here, the accompanying \s--flows in
the  string  equation  \smileyii\  are  studied.  The  solutions were found
primarily  using  the  FORTRAN  NAG  finite  element  library  routine {\sl
D02RAF}.  This routine  allows the  solution of  several simultaneous first
order ordinary differential equations with given boundary conditions by use
of Newton  interpolation from an  initial approximate solution  supplied by
the  user.
The  equation $\qutr\R'''-\hf\rho^{'}\R-(\rho-\sigma)\R^{'}=0$  was
studied, which is  the first differential of \smileyii\  with a factor $\R$
removed.  This enabled  better  calculation  of the  numerical derivatives:
solving  \smileyii\  directly  would  involve  the  implicit calculation of
$1/\R$, which diverges for large positive $z$, as $\R=0$ in that asymptotic
limit.  The above  equation, a  differential equation  of order $2m+1$, was
transformed to $2m+1$ first order equations. Boundary conditions were taken
from the above asymptotics.

The first case studied in this way  was $m=1$. The equation was solved with
$\R={3\over  2}\rho-z$ on  a mesh   of 2000  points. The  routine converged
rapidly to the solution for each value  of \s\ with an error of better than
$10^{-4}$.  The results  are shown  in figure  \fig\mone{The \s-flow in the
$m=1$ model for positive and negative \s.}. The
solution for $m=2$ with $\sigma$ positive  was carried out in the same way.
Here $\R={5\over 2}(-\third\rho^{''}+\rho^2)-z$, and  a mesh of 2000 points
was used. The  results are displayed in figure  \fig\mtwoa{The \s-flow in the
$m=2$ model (pure gravity) for positive  \s.}.
For  $\sigma$ negative,  however, the  solution becomes  progressively more
oscillatory,  and  the  derivatives  take  on  increasingly  large  values.
Eventually  the required  accuracy could  no longer  be obtained using {\sl
D02RAF}. Here, the \s--flow equation was employed to make further progress.
In  order  to  solve  for  more  negative  $\sigma$,  the solution for some
previous  value of  $\sigma$ is  used as  a starting  solution. A numerical
integration of equation\sigflow\ is then  carried out using Euler's method.
This gives the next initial approximation for {\sl D02RAF}, and the process
can be iterated  to give solutions for more  negative $\sigma$. The results
are  shown  in  figure  \fig\mtwob{The \s-flow in the
$m=2$ model (pure gravity) for negative  \s.}.  Unfortunately,  at
$\sigma=-1.25$, both rounding errors and  systematic errors become so large
(of  order  1,  and  much  larger  for  higher  derivatives)  as to prevent
convergence. Rounding errors, and the reliance on high derivatives, prevent
the use of more accurate  integration methods such as Runge-Kutta. However,
the pattern of behaviour of the  solution for increasingly negative \s\ has
been established.  The $m=3$ solution  was found in  a straightforward way,
using  the method  described for  the $m=2$,  $\sigma$ negative case above.
Here  however, there  were no  problems with  convergence, and  the program
required relatively  few mesh points  to give small  errors (typically 2000
for  error  of  $<  10^{-4}$,  against   15,000  for  the  $m=2$  case  for
$\sigma<0$). Results are shown in figure \fig\mthree{The \s-flow in the
$m=3$ model for positive and negative  \s.}.

When studying these curves, the  associated double scaled Dyson gas problem
should be recalled\ref\multi{S. Dalley, C. Johnson and T. Morris, \NP{B368}
(1992)  625.}\ref\noflow{M.R.Douglas, N.Seiberg  and S.Shenker,  \PL{B244}\
(1990)  381.}\npqg\npqga\simon. In  scaled coordinates  $\lambda_s$, in the
spherical  approximation,  the  charges  are  concentrated  on a single cut
extending  along  the  semi--open  interval  $[\rho,\infty)$.  There  is an
infinite  potential wall  at position  $\lambda_s=\sigma$, restricting  the
problem to $\rline_+$.

Without  the wall,  the topology  is $\rline$  which leads  to the critical
physics  of  the  \pq  definition.  The  critical  physics  of the model is
entirely characterised by the local behaviour  of the charge density in the
neighbourhood of  its endpoint\ref\class{S.Dalley, C.Johnson  and T.Morris,
\MPL{A6}  (1991),  439.},  at  position  $\lambda_s=\rho$.  The  density of
charges vanishes  at least as  a square root  at its endpoint,  and for the
$m$th critical model  it has $m-1$ extra zeros  there. The resulting string
equation derived  for $\rho$ is $\R=0$.  In examining the stability  of the
critical models a  study at the spherical level  of the effective potential
for  one  eigenvalue  reveals  that  for  an  odd  number  of  zeros  it is
energetically favourable for  eigenvalues to leave the single  cut and move
to  a new  configuration\simon\npqg. This  is the  origin of  the $m$--even
models' instability.

In  the  \pqq  formulation  the   effective  potential  described  above  is
supplemented  by  the  infinite  potential  wall,  which  has the effect of
stabilising  the problem\npqg.  In the  asymptotic analysis  at large  $z$,
there is the following behaviour:  For $z\to+\infty$ the density pulls away
from the wall, and the neighbourhood of the endpoint is the same as that of
the \pq definition. Therefore the genus by genus physics for large positive
$z$  is identical  to that  of the  \pq definition.  At the string equation
level,  this  is  realised  by  the  fact  that  the  solution  of equation
\smileyii\ satisfies  $\R=0$ in this  limit. For $z\to-\infty$  the density
pushes up  against the wall,  and the  square  root zero is  replaced by an
integrable  square root  divergence. This  leads to  $\rho=\sigma$ for  the
leading behaviour of the endpoint.

The  non--perturbative  positions  of  the  charges  in  the  Dyson gas are
directly  related  to  the  spectrum  of  the  hamiltonian  ${\cal H}\equiv
Q=\nu^2\partial_z^2-\rho$. For the zero \s\  case, the single cut nature of
the charge density is consistent with a continuous spectrum for \h\ with no
discrete states  below the continuum.  This may be  confirmed by examining,
for example, the form of the solution  for $\rho$ for $m=1$. The small well
is too  shallow to support such  discrete states. This is  immediately true
for all  of the $m$--critical solutions  for $\rho$: they are  connected by
global  KdV  flows\Global\  and  the  KdV  flows  are the {\sl isospectral}
deformations of the hamiltonian \h.

Moving on to  discuss the non--zero \s\ case, we  first consider the nature
of  the  \s--flows.  They  represent  the  first  in  a series of important
{non--isospectral}  deformations of  \h, the  Galilean transformations. The
next  in this  series are  the scalings.  Demanding invariance  under these
transformations leads  to the $L_{-1}$ and  $L_0$ Virasoro constraints. The
rest   of  the   Virasoro  constraints   follow  from   higher,  non--local
transformations which may  be derived using the recursion  operator for the
KdV hierarchy.  Using Galilean transformations, the  continuous spectrum of
\h\  cannot be  deformed to  a spectrum  with any  discrete states. This is
consistent with the Dyson gas  picture where the Virasoro constraints arise
from diffeomorphisms in $\rline_+$, with the  terms due to \s\ arising from
varying the  position of the wall.  By definition, diffeomorphisms preserve
the  single  cut  nature  of  the  charge  density,  and  thus  the bounded
continuous spectrum of \h. In particular,  $L_{-1}$ which is the \s-flow at
the $\tau$-function level, is a translation of the Dyson gas and hence just
a shift of the spectrum of \h.

The  case of  $m=1$ is  particularly simple.  Starting with $\sigma=0$, the
Galilean transformation \galileo\ may be used  to move to non--zero \s. The
parameter $t_1$ remains  invariant, and it may be set  to 1 without loss of
generality.  The result  of this  process at  some finite  \s\ (positive or
negative)  is  a  shift  in  $\rho$  and  $z$  by  \s.  This corresponds to
translating the  whole $\sigma=0$ curve  along the diagonal  line $\rho=z$.
This is the  process illustrated in figure~\mone, which  is the solution of
\smileyii\  at $m=1$  for arbitrary  values  of  \s. Away  from $m=1$,  the
processes  involved  are  not  as  simple.  This  is  because  the Galilean
transformation inevitably switches on  couplings $t_k$ ($k<m$) to operators
other than the  puncture operator $\O_0$, $\O_m$ and  $\O_B$. For \s\ going
positive at  fixed $z$ the wall  moves to the right,  in scaled coordinates
$\lambda_s$, and pushes up towards the  density. This is the reverse of the
$z\to-\infty$  limit  for  fixed  \s\  discussed  before and asymptotically
$\rho=\sigma-\nu^2/(4z^2)+\ldots$.  In this  large \s\  regime, surfaces of
finite area are  increasingly suppressed and the model  is dominated by the
$m=0$  topological phase.  See figures~\mone~\mtwoa\  and \mthree.  For \s\
increasingly negative,  the wall moves  to the left,  and the solution  for
$\rho$  gradually  begins  to  resemble  the  \pq  $m$--critical  solution,
although this process can never be  completed for finite \s. In particular,
for  the $m=2$  (pure gravity)  case, the  solution develops  progressively
sharper oscillations for negative sigma. Naively, this suggests that in the
limit  $\sigma\to-\infty$, the  oscillations evolve  into the  poles of the
Painlev\'e~I equation.  However, this is  an ill--defined limit,  as can be
seen from  many points of view.  There is no way  to continuously deform to
such a solution  of Painlev\'e~I in the same sense  that it is not possible
to connect  the BMP solution  for the \pq  definition of
 $m=3$ to  it via the KdV
flows\noflow. Furthermore the spectrum of \h\ for the singular solutions of
Painlev\'e~I  is discrete,  and there  is no  way to  deform the continuous
spectrum  of  \h\  for  solutions  of  \smileyii\  into  a discrete one via
\s-flow,  as discussed  above. This   generalises to  all of  the $m$--even
cases. For $m$--odd, the spectrum is  continuous and not bounded from below
for the  BMP--type solutions of the  \pq definition, and it  is conceivable
that   they  are   connected  to   the  \pqq   $m$--odd  solutions  in  the
$\sigma\to-\infty$ limit.  This picture is possibly  consistent with one in
which the wall has moved away sufficiently far  to the left so as to have a
vanishing effect on the non--perturbative physics. For the $m$-even models,
the reduced presence of the wall in this limit allows the non--perturbative
physics to  more resemble that of  the \pq definition, but  it can never be
completely removed so as to recover instability.

\bigskip
\bigskip
\noindent
{\bf Acknowledgements}

\noindent
It is a pleasure to thank Simon Dalley for helpful comments.
C.J. thanks the S.E.R.C. for
financial support.

\listrefs
\listfigs
\bye